\documentclass[aps,prl,superscriptaddress,floatfix,twocolumn,longbibliography]{revtex4-2}

\usepackage{amsmath,dcolumn,bm,amsthm,tabularx,subfigure}
\usepackage[colorlinks=true,urlcolor=blue,citecolor=blue,linkcolor=blue]{hyperref}
\usepackage{mathrsfs,times}
\usepackage{bbold,dsfont}
\usepackage{graphicx,graphics,color}
\usepackage{mathtools,nicefrac}
\usepackage{slashed}
\usepackage{txfonts}
\usepackage{amsfonts,amssymb}
\allowdisplaybreaks

\newcommand{\beq}{\begin{equation}}
\newcommand{\eeq}{\end{equation}}
\newcommand{\bea}{\begin{eqnarray}}
\newcommand{\eea}{\end{eqnarray}}

\newcommand{\XX}{{XX}}
\newcommand{\ZZ}{{ZZ}}
\newcommand{\XZ}{{XZ}}
\newcommand{\hog}{{\big|\frac{h}{g}\big|}}

\begin{document}

\title{Superdiffusive to Ballistic Transports in Nonintegrable Rydberg Chains}

\author{Chun~Chen}
\email[Corresponding author.\\]{chunchen@sjtu.edu.cn}
\affiliation{Key Laboratory of Artificial Structures and Quantum Control (Ministry of Education), School of Physics and Astronomy, Shenyang National Laboratory for Materials Science, Shanghai Jiao Tong University, Shanghai 200240, China}

\author{Yan~Chen}
\email[Corresponding author.\\]{yanchen99@fudan.edu.cn}
\affiliation{Department of Physics and State Key Laboratory of Surface Physics, Fudan University, Shanghai 200433, China}

\author{Xiaoqun~Wang}
\email[Corresponding author.\\]{xiaoqunwang@zju.edu.cn}
\affiliation{Key Laboratory of Artificial Structures and Quantum Control (Ministry of Education), School of Physics and Astronomy, Shenyang National Laboratory for Materials Science, Shanghai Jiao Tong University, Shanghai 200240, China}
\affiliation{School of Physics, Zhejiang University, Hangzhou 310058, Zhejiang, China}
\affiliation{Tsung-Dao Lee Institute, Shanghai Jiao Tong University, Shanghai 200240, China}
\affiliation{Collaborative Innovation Center of Advanced Microstructures, Nanjing University, Nanjing 210093, China}

\date{\today}

\begin{abstract}

A common wisdom posits that transports of conserved quantities across clean nonintegrable quantum systems at high temperatures are diffusive when probed from the emergent hydrodynamic regime. We show that this empirical paradigm may alter if the strong interaction limit is taken. Using Krylov-typicality and purification matrix-product-state methods, we establish the following observations for the strongly interacting version of the mixed-field Ising chain, a nonintegrable lattice model imitating the experimental Rydberg blockade array. Given the strict projection owing to the infinite density-density repulsion $V$, the chain's energy transport in the presence of a transverse field $g$ is superdiffusive at infinite temperature featured by an anomalous scaling exponent $\frac{3}{4}$, indicating the existence of a novel dynamical universality class. Imposing, in addition, a growing longitudinal field $h$ causes a drastic factorization of the whole Hilbert space into smaller subsectors, evidenced by the spectral parsing of the eigenstate entanglement. Being a consequence of this approximate symmetry, a superdiffusion-to-ballistic transport transition arises at $h\approx g$. Interestingly, all the above results persist for large but finite interactions and temperatures, provided that the strongly interacting condition $g,h\ll k_\textrm{B}T\ll V$ is fulfilled. Our predictions are verifiable by current experimental facilities.       

\end{abstract}

\maketitle

{\it {\color{blue}Introduction.}}---In quantum dynamics, states of matter can be classified by their universal transport properties, e.g., normal diffusion for nonintegrable systems and ballistic propagation for integrable models \cite{ZotosReview,Polkovnikov,Bertini}. In this regard, one question central to nonequilibrium many-body physics, quantum simulation, and statistical mechanics concerns the modifications to this standard paradigm stemming from the interplay among distinct energy scales of the problem. For closed setups, there are $3$ such quantities, the free and the interaction parts of the Hamiltonian and the temperature. Most previous literature focused on the influence of the moderate interactions on the system's low-temperature transport behaviors \cite{Zotos,Sirker}. Intriguingly, a recent advance in this context takes a different angle that leads to the finding of the Kardar-Parisi-Zhang (KPZ) \cite{KPZ} spin superdiffusion at infinite temperature in the integrable Heisenberg chain with a fine-tuned non-Abelian symmetry \cite{Znidaric,Ljubotina,Scheie,WeiBloch}. It is noteworthy that here temperature sets the maximum scale that is far beyond the scale of the Hamiltonian \cite{Dupont,KimHuse}.

In this Letter, we consider the scenario where the interaction part of the Hamiltonian $H_{\textrm{int}}$ is first taken to infinity and the subsequent measurements and evaluations are then performed in the infinite-temperature limit while the free part of the Hamiltonian $H_0$ is maintained as the reference point. Physically, such an arrangement corresponds to the strongly interacting condition, $H_0\ll k_\textrm{B}T\ll H_{\textrm{int}}$. Here we are chiefly inspired by the recent progress on the Rydberg blockade experiments \cite{Bernien}, where the above construction appears to be achievable. The anomalous energy transports in the clean Rydberg blockade chain, as will be reported in detail in the current work, were first hinted in Refs.~\cite{ChenChenWang,ChenChenWangLR} where we extensively discussed the randomness-induced dynamical effects in Rydberg array when subject to this strong interaction limit.      

{\it {\color{blue}Clean constrained model.}}---Neutral atoms in optical tweezers with large principal quantum numbers can be photon driven into Rydberg qubits. The generated electric dipole-dipole interactions entangle atoms within a blockade sphere, accommodating no more than one Rydberg excitation. Lasers are then used to address individual atoms for the onsite qubit operations. Under the correspondence $b_i^\dagger+b_i=\sigma_i^x,\ b_i^\dagger b_i=n_i=\frac{1}{2}(1-\sigma_i^z)$ upon basis $|{\sf g}\rangle=|\!\!\uparrow\rangle,\ |{\sf r}\rangle=|\!\!\downarrow\rangle$, one canonical hard-core boson model realizable in such Rydberg analog simulators imitates the nonintegrable quantum Ising chain in the transverse and longitudinal fields \cite{Bernien},
\beq
H=\sum_i[g(b^\dagger_i+b_i)+h(1-2n_i)+V n_i n_{i+1}], \label{constrhamboson}
\eeq
where Rydberg blockade means $V=\infty$, then specifically the constrained model \cite{Chen} reads
\beq
\widetilde{H}=\sum_i[gP(b^\dagger_i+b_i)P+hP(1-2n_i)P] \label{constrhamboson_proj}
\eeq
with projector $P=\prod_i(1-n_in_{i+1})$, thereby preventing $2$ simultaneous excitations for any pair of neighboring atoms. Rabi frequency and detuning of the coherent laser beam are designated by the uniform transverse and longitudinal field strengths $g$ and $h$, respectively. Throughout this paper, we use hard-core boson and spin-$\frac{1}{2}$ representations interchangeably, choose $g=1$ as the energy unit, and keep $\hbar=k_\textrm{B}=a=1$.

\begin{figure}[t]
\centering
\includegraphics[width=1\linewidth]{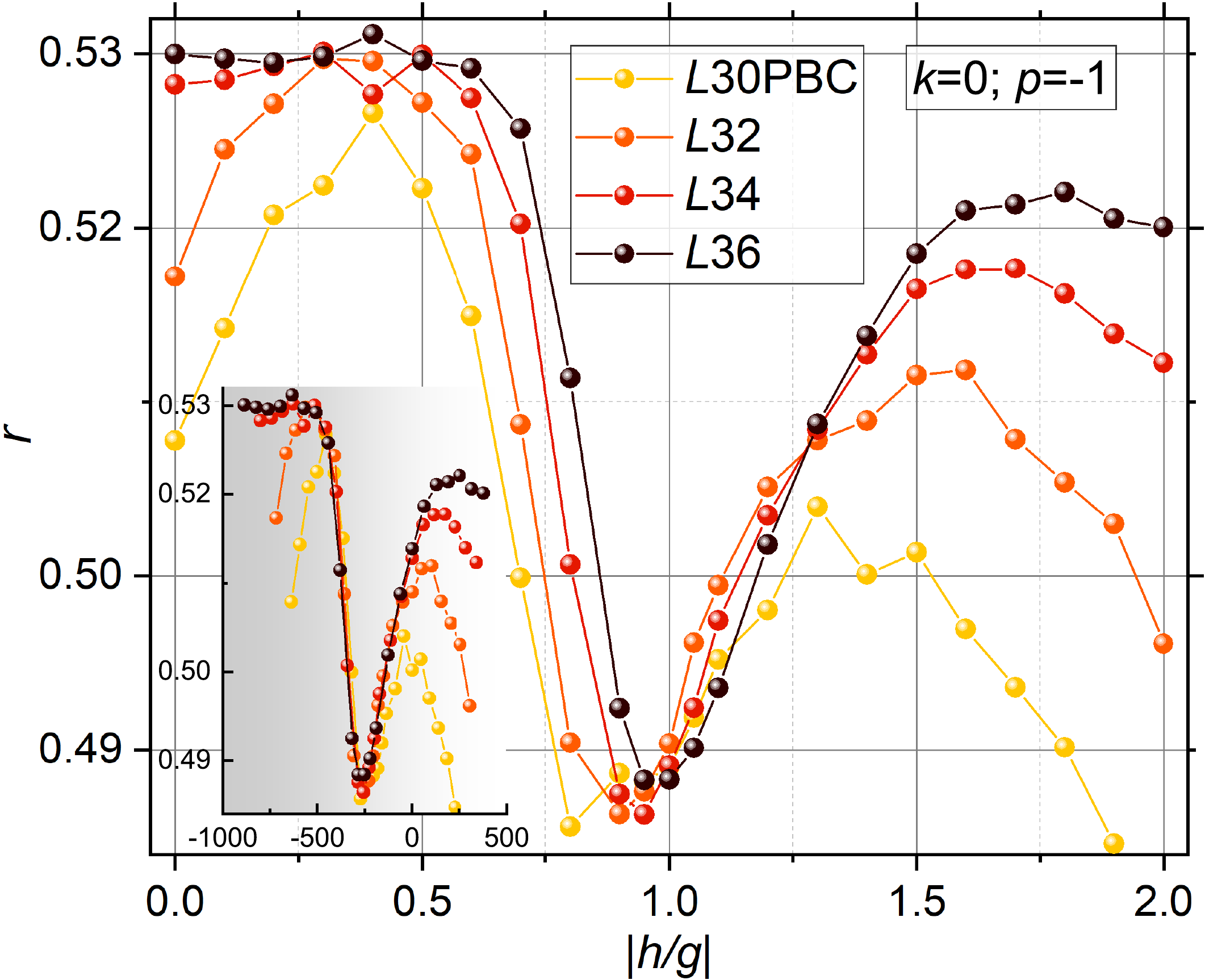}
\caption{\label{fig:fig1} Finite-size scaling of mean level-spacing ratio $r$ as a function of $\hog$ in the zero-momentum, odd-parity $(k=0,p=-1)$ sector of the clean periodic Rydberg blockade chain [model (\ref{constrhamboson_proj})] up to maximum length $L=36$. The inset illustrates the accompanying data collapse after adjusting the horizontal axis to $(h-1.4)L^{1.8}/g$ while maintaining the vertical axis to $r$, wherein a dip of $r$, dropping below $0.49$, is surrounded by two GOE plateaus.} 
\end{figure}

{\it {\color{blue}Static spectral \& entanglement diagnostics.}}---Because time-reversal symmetry is preserved, for generic nonzero parameters, clean mixed-field Ising chain including its infinitely interacting version is anticipated to obey eigenstate thermalization hypothesis (ETH) \cite{Deutsch,Srednicki} whose eigenspectrum shall be describable by Gaussian orthogonal ensemble (GOE) \cite{DAlessio}. Using exact diagonalization, we scrutinize the spectral properties of the blockade model (\ref{constrhamboson_proj}) via computing its level statistics. The level-spacing ratio is defined by $r_n=\frac{\min\{\delta_n,\delta_{n-1}\}}{\max\{\delta_n,\delta_{n-1}\}}$ whose mean $r\approx0.53$ if GOE is assumed where $\delta_n=E_n-E_{n-1}$ with $\{E_n\}$ an ascending set of all available eigenvalues \cite{Oganesyan}.

Figure~\ref{fig:fig1} shows the evolution of the mean ratio $r$ under the increase of $\hog$ for the periodic Rydberg chain in the blockade regime. By exploiting translation and reflection symmetries, we pick the zero-momentum, odd-parity sector of the model and push the maximum system size to $L=36$. The calculated spectral statistics exhibit several peculiarities. (i) Within $0\leqslant\hog\lesssim1$ (relative sign between $g,h$ is immaterial), the system's energy-level distribution is well captured by GOE. (ii) However, once beyond $\hog\gtrsim1.5$, this trend slows down and longer chains are entailed to witness the full convergence toward GOE further. (iii) In between, these two GOE plateaus are separated by a sudden dip where $r$ deviates from $0.53$ appreciably and drops below $0.49$, indicating a potential eigenstate phase transition between these two thermal regions. As illustrated by the inset of Fig.~\ref{fig:fig1}, the companion finite-size data collapse of $r$ supports the finding of the transition where a critical $\hog\approx1.4$ can be identified.

Consequently, what are the fundamental distinctions between the thermal phases at small and large $\hog$? To this end, we proceed to investigate the energy-resolved entanglement entropy for the entirety of the eigenstates of model (\ref{constrhamboson_proj}). For each eigenfunction with eigenenergy $E$, we compute its half-chain von Neumann entropy $S_{\textrm{vN}}=-{\sf Tr}\rho_R\log_2\rho_R$ via reduced density matrix $\rho_R$ by tracing out the degrees of freedom on left half-chain and then plot $S_{\textrm{vN}}$ as a function of $E$.

\begin{figure}[t]
\centering
\includegraphics[width=1\linewidth]{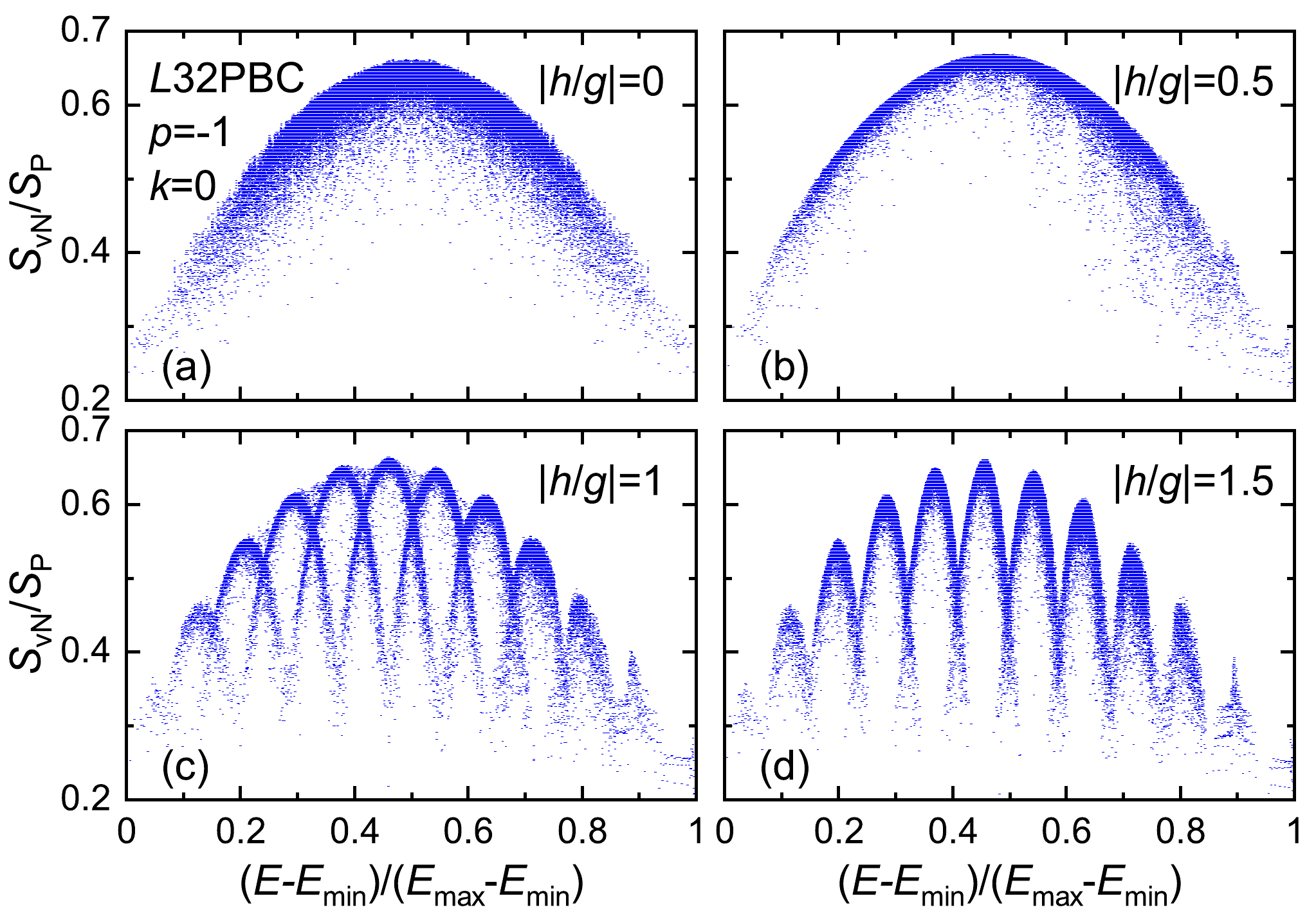}
\caption{\label{fig:fig2} Spectral structure of entanglement entropy for periodic Rydberg blockade chain of length $L=32$ within zero-momentum, odd-parity sector. For every eigenstate, a point is drawn whose coordinates are marked by its eigenenergy and entanglement entropy. (a)-(d) correspond to $\hog=0,0.5,1,1.5$, respectively.} 
\end{figure}

Figure~\ref{fig:fig2} depicts the spectral parsing of $S_\textrm{vN}$ for the periodic Rydberg blockade chain at $4$ representative parametric points across regions of small, intermediate, to large values of $\hog$. In accord with the signature of a transition in Fig.~\ref{fig:fig1}, the distribution of $S_\textrm{vN}$ along the eigenenergy axis displays $2$ distinctive patterns. As shown by Figs.~\ref{fig:fig2}(a),(b), for thermal phase at small $0\leqslant\hog<1$, the entanglement data tend to form a single, smoothing curve following the prediction of ETH \cite{Turner}. This tendency alters dramatically once $|h|$ exceeds $|g|$. Figs.~\ref{fig:fig2}(c),(d) demonstrate that in this circumstance, the whole Hilbert space of the constrained chain appears to factorize into smaller subsectors, each resembling the line shapes of Figs.~\ref{fig:fig2}(a),(b) but with negligible interference.

Physically, these subsectors emerging once $\hog\gtrsim1$ can be roughly tagged by the total number of hard-core bosons $N_b=\sum_in_i$ in the Rydberg chain. Besides the constraint, the selection of the allowed values for $N_b$ is also subject to the compatibility condition with the additional symmetry indices, for instance, the momentum and/or parity quantum numbers. If without using auxiliary symmetries, there typically exist about $\frac{L}{2}$ such subsectors corresponding to the $\frac{L}{2}$ possible values of $N_b$ for a blockade chain of length $L$. Since $|g|$ is in the same order as $|h|$, $N_b$ is not a good quantum number. Moreover, it is noticeable from Fig.~\ref{fig:fig1} that the model's level statistics within this parameter region approach GOE. In this regard, such an approximate symmetry of $N_b$ steered by the longitudinal field $|h|$ is not expected, which probably implies a hidden constraint-facilitated effectiveness of $|h|$ in suppressing the inter-subsector mixture induced by the transverse flip term $|g|$. As the total number of subsectors is proportional to $L$, the above observations of Fig.~\ref{fig:fig2} are different from the Hilbert-space fragmentation where this growth is an exponential function of $L$ \cite{SalaPollmann,KhemaniHermele}.

\begin{figure*}[tb]
\centering
\includegraphics[width=0.996\linewidth]{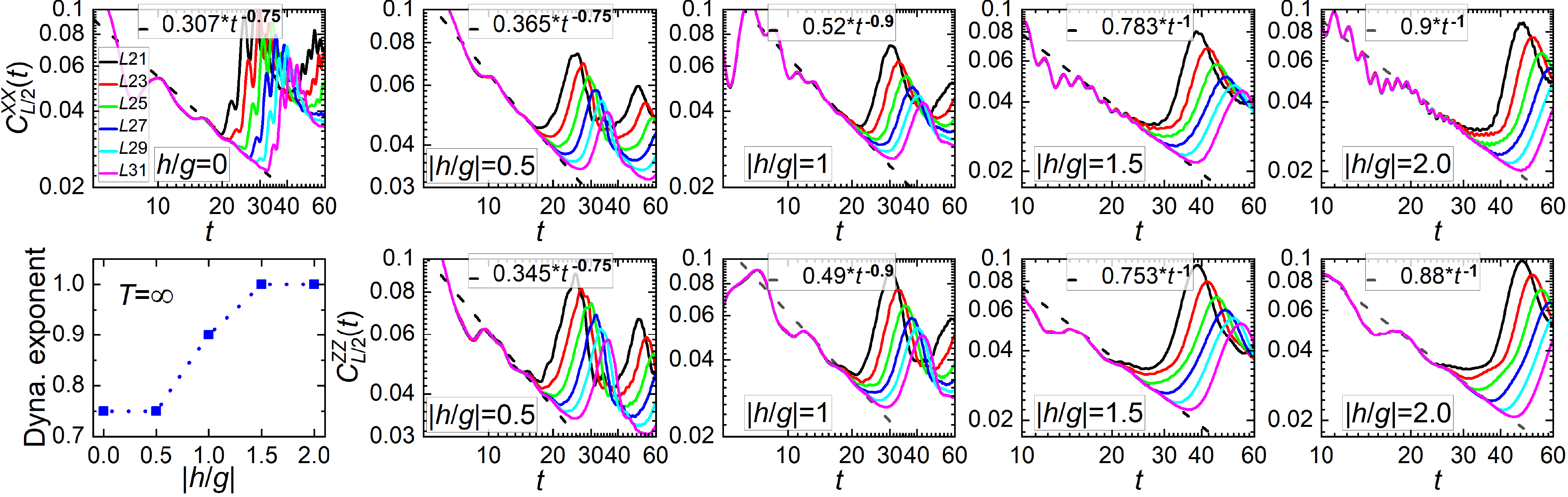}
\caption{\label{fig:fig3} Time evolution of the return probability at infinite temperature using OBCs. The upper (lower) row addresses the $\XX\ (\ZZ)$ component. For model (\ref{constrhamboson_proj}) at $T=\infty$, the $\XZ$ component vanishes. A range of parameters $\hog=0,0.5,1,1.5,2$ is organized into $5$ columns. The Krylov-typicality technique solves system sizes ranging from $L=21$ to $31$. The results converge onto a power law of time, whose exponent from fitting is summarized in the bottom-left panel as a function of $\hog$.} 
\end{figure*}

{\it {\color{blue}Dynamical scaling exponent from return probability.}}---Contrast between thermal phases at small and large $\hog$ becomes reinforced when inspecting how conserved energy is transported across the blockade chain under the unitary time evolution. At high temperatures, transportation of conversed quantity in generic nonintegrable isolated systems is obedient to the diffusive behavior, which is particularly commonplace for energy propagation.

We find that this paradigm fails in Rydberg blockade chain. To examine the transport properties of model (\ref{constrhamboson_proj}) where energy is the only conserved quantity, we utilize Krylov-typicality technique \cite{Steinigeweg,Paeckel} to numerically evaluate the connected equal-site time-dependent return probability defined by \cite{KimHuse},
\beq
C^{\mathcal{OO}}_{L/2,\beta}(t)=\langle \delta\rho_{\mathcal{O}}(L/2,t)\delta\mathcal{O}(L/2,0)\rangle\propto t^{-\alpha},
\label{returnprob}
\eeq  
where $\langle\cdots\rangle\coloneqq{\sf Tr}[e^{-\beta \widetilde{H}}\cdots]/{\sf Tr}[e^{-\beta\widetilde{H}}]$, $\rho_{\mathcal{O}}(L/2,t)=e^{i\widetilde{H}t}P_{\frac{L}{2}-1}\mathcal{O}(L/2)P_{\frac{L}{2}+1}e^{-i\widetilde{H}t}$ with $\delta\rho_{\mathcal{O}}(L/2,t)\coloneqq\rho_{\mathcal{O}}(L/2,t)-\langle\rho_{\mathcal{O}}(L/2,t)\rangle$ is the density matrix associated to the component $\mathcal{O}$ of the Hamiltonian density on site $\frac{L}{2}$, $P_i=1-n_i$ is the local projector to render the initial energy disturbance concentrated around the chain center thus allowing for a convenient inspection on how this inhomogeneity gets smeared, and $\beta=\frac{1}{k_\textrm{B}T}$ is the inverse temperature. At long wavelengths, hydrodynamic fluctuations dominate the collective motion involving many particles, giving rise to a universal power-law scaling $t^{-\alpha}$ for the later-time dependence of return probability. This dynamical scaling exponent $\alpha$ provides a scheme to classify the transport behaviors into diffusive when $\alpha=\frac{1}{2}$ or ballistic when $\alpha=1$. In between, the transport is subdiffusive if $0<\alpha<\frac{1}{2}$ or superdiffusive if $\frac{1}{2}<\alpha<1$.

In Fig.~\ref{fig:fig3}, we focus attention on $C^{\mathcal{OO}}_{L/2}(t)$ at infinite temperature $(\beta=0)$ and choose $\mathcal{O}=\widetilde{X},\widetilde{Z}=P\sigma^{x,z}P$. Deploying the Krylov-typicality approximation by averaging over $1000$ independent pure states, we evolve a spectrum of open chains with length from $L=21$ up to $31$. After the initial transient oscillations, all curves of different sizes converge to one steady power-law decay of time. Intriguingly, rather than diffusion, the scaling exponent of the energy transport throughout the small-$\hog$ regime is close to $\frac{3}{4}$, indicating a superdiffusion. As no long-range interactions exist, L\'evy flight \cite{Zaburdaev} is not immediately relevant. Moreover, this exponent of value $\frac{3}{4}$ does not belong to the Fibonacci family of dynamical universality classes \cite{Popkov}, of which the Gaussian diffusion with exponent $\frac{1}{2}$ and the KPZ superdiffusion with exponent $\frac{2}{3}$ are $2$ members.

By contrast, the large-$\hog$ regime is featured by a ballistic energy transport typical for integrable systems. On the one hand, this ballistic behavior is unexpected because from Fig.~\ref{fig:fig1} the level statistics inside this thermal phase exhibit a mounting tendency toward GOE, thereby being nonintegrable and shall predict a diffusive behavior. On the other hand, an indication of the emergent approximate $N_b$ symmetry evidenced by Fig.~\ref{fig:fig2} hints at the consistency between this ballistic transport and the quasi-integrals of motion associated to $N_b$. However, although the number of the permitted values of $N_b$ is linear in $L$, it is far from exponentially many compared to the Hilbert-space dimension. This partly explains why, in large-$\hog$ region, $r$ is GOE rather than Poisson. Accordingly, reasonings assumed integrability may not be applicable to the case at hand.

\begin{figure*}[t]
\centering
\includegraphics[width=0.996\linewidth]{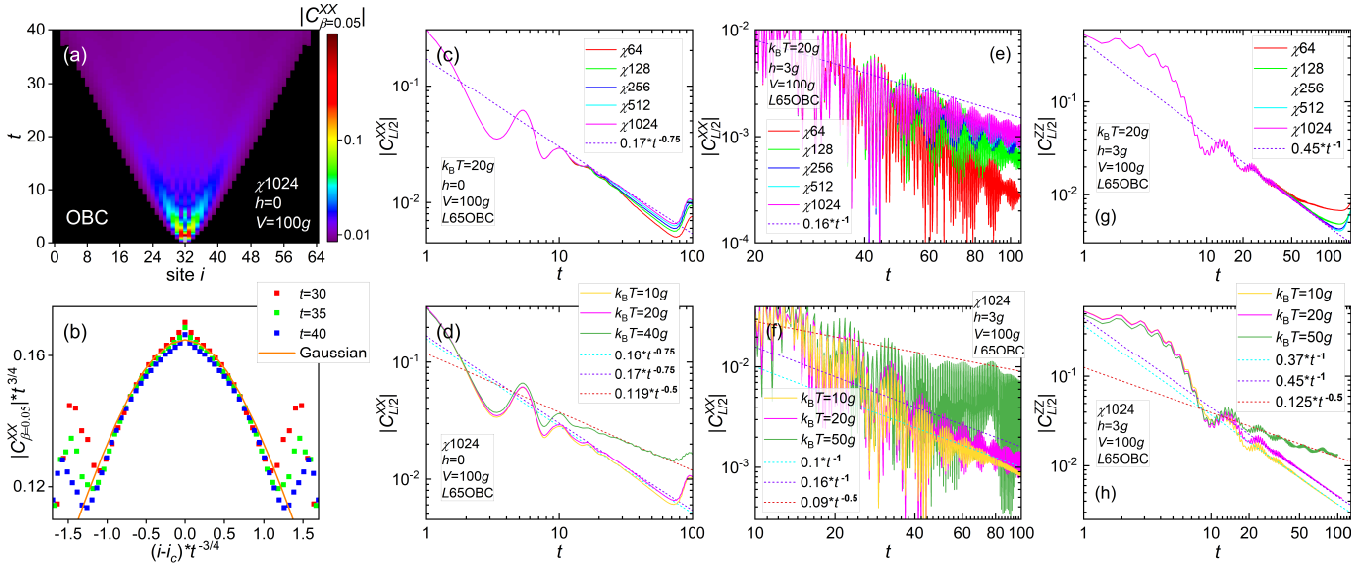}
\caption{\label{fig:fig4} Purified TEBD results for the finite-$V$ and finite-$T$ energy transports of model (\ref{constrhamboson}) using an open Rydberg chain of length $L=65$. Panels (a)-(d) focus on demonstrating the anomalous superdiffusion in the pure transverse field limit $(\frac{h}{g}=0)$, while panels (e)-(h) switch to illustrate the realization of a ballistic energy transport after enhancing the longitudinal field strength to $h=3g$. For more explanations, see text.} 
\end{figure*} 

The proposed eigenstate transition between small- and large-$\hog$ phases is now signaled by a crossover between superdiffusive to ballistic energy transports. In Fig.~\ref{fig:fig3} we show that close to criticality the scaling exponent $\alpha\approx0.9$ when $\hog=1$. Physically, the $x,z$ quantization axes are different due to the infinite-interaction projection. But Fig.~\ref{fig:fig3} reveals that their respective return probabilities are nevertheless governed by the same scaling exponent, suggesting energy transport is likely isotropic in the spin space, consistent with the expectations from GOE and ETH. Furthermore, we check that the entanglement spreadings in both thermal regions are ballistic, similar to the conventional ETH phases \cite{KimHuse}. Therefore, the computational resource for capturing the long-time dynamics of longer Rydberg chains should grow promptly.

{\it {\color{blue}Finite-$V$ \& finite-$T$ results.}}---We have exclusively focused on model (\ref{constrhamboson_proj}) which is derived from the more realistic model (\ref{constrhamboson}) by first sending $V\rightarrow\infty$ and then $T\rightarrow\infty$. Now we switch to the purification TEBD method \cite{Vidal,Schollwock,Karrasch} to show that the energy-transport properties found so far are quantitatively preserved in the model (\ref{constrhamboson}) at finite $V$ and $T$, thus being experimentally measurable. The essence is to achieve the large-$V$ and high-$T$ condition, $g,h\ll k_\textrm{B}T\ll V$. To this end, we fix $V=100g$ and tune $h\in[0,3g]$ and $k_\textrm{B}T\in[10g,50g]$ in the following matrix-product-state simulation.

Figures~\ref{fig:fig4}(a)-(d) target the case of superdiffusion at $h=0$. Via computing the $\XX$ component of the return probability at $k_\textrm{B}T=20g$ by replacing $\widetilde{H}$ with $H$ and setting $\mathcal{O}=\sigma^{x}$ in Eq.~(\ref{returnprob}), we illustrate in Fig.~\ref{fig:fig4}(c) up to the maximum bond dimension $\chi=1024$ that in accord with the top-left panel of Fig.~\ref{fig:fig3}, the same type of energy superdiffusion characterized by a fitted dynamical exponent $\frac{3}{4}$ arises also on a longer open Rydberg chain of length $L=65$ described by model (\ref{constrhamboson}). (Finite-size analyses covering $L=33,49,65$ yield the same value of $\alpha$.) Furthermore, Fig.~\ref{fig:fig4}(d) unveils that with fixed $\chi=1024$, the just-claimed superdiffusion occurs at $k_\textrm{B}T=10g$ as well. Thus, a superdiffusive phase is stabilized in a temperature range of $k_\textrm{B}T\in[10g,20g]$. Increasing the temperature to $k_\textrm{B}T=40g$, however, restores the system's transport to a normal diffusion characterized by $\alpha=\frac{1}{2}$. In this sense, a super-to-normal diffusion crossover exists in the model (\ref{constrhamboson}) triggered solely by temperature.

The dynamical universality class for the anomalous transport phenomenon is dictated by the scaling exponent and the scaling function. By changing the $2$nd operator position in Eq.~(\ref{returnprob}) from the chain center to the end, we plot the corresponding spatiotemporal evolution contour of the connected correlation function $|C^{\XX}_{\beta=0.05}|$ in Fig.~\ref{fig:fig4}(a). Physically, it is reasonable to assume that the correlation function in the hydrodynamic regime satisfies the scaling relation $|C^{\XX}_{\beta=0.05}|\propto t^{-\alpha}f[(i-i_c)/t^{\alpha}]$ where $\alpha$ is the scaling exponent and the unknown function $f[\cdots]$ is the scaling function. Although the system sizes and the time scales achieved in the present work do not allow for a reliable investigation of the detailed form of the scaling function, in Fig.~\ref{fig:fig4}(b), we proceed to show the preliminary data collapse of the $|C^{\XX}_{\beta=0.05}|$ profiles collected at $3$ different moments from Fig.~\ref{fig:fig4}(a) and compare it to a Gaussian fit. With $\alpha=\frac{3}{4}$ fixed, we find that these various data points do fall onto a single curve at large $i,t$, and this line decays rapidly, thus becomes progressively deviating from the Gaussian fit. This trend is consistent with the prediction of the superdiffusive behavior by the scaling exponent $\frac{3}{4}$.

Figures~\ref{fig:fig4}(e)-(h) target the case of ballistic energy transport at $h=3g$. The signature for this ballistic diffusion is most transparent from Fig.~\ref{fig:fig4}(g) where the $\ZZ$ component of the return probability exhibits a clean convergent tendency toward the characteristic linear power-law decay with scaling exponent $1$ at $k_\textrm{B}T=20g$ under the enhancement of $\chi$. Further, despite the extra complication from the more significant oscillations, the accompanying $\XX$ component of the return probability converges to the same ballistic transport as displayed by Fig.~\ref{fig:fig4}(e). Therefore, threading a growing longitudinal field while maintaining other parameters intact induces a field-tuned crossover between the anomalous superdiffusion at small $h$ [Fig.~\ref{fig:fig4}(c)] and the ballistic diffusion at moderate $h$ [Figs.~\ref{fig:fig4}(e),(g)]. In addition, by varying $k_\textrm{B}T$ at fixed $\chi=1024$, Figs.~\ref{fig:fig4}(f),(h) show that this ballistic behavior persists down to $k_\textrm{B}T=10g$. Conversely, lifting the temperature to $k_\textrm{B}T=50g$ turns the system's energy propagation to the normal diffusion. Accordingly, at medium $h$, a temperature-tuned crossover exists from the ballistic to the diffusive energy transport.         

{\it {\color{blue}Conclusion \& outlook.}}---Based on large-scale numerical studies of $1$D Rydberg blockade array, we predict that under strong interaction conditions, there exist two distinct types of constrained thermal phases. The first arises in the transverse-field dominated regime whose energy transport is superdiffusive featured by a dynamical exponent $\frac{3}{4}$. The second is stabilized after adding a growing longitudinal field whose energy diffusion is ballistic. Complementary spectral and dynamical analyses suggest that there might exist an eigenstate transition between the two constrained thermal phases. An analytical understanding of these numerical findings is lacking. Future works can be oriented toward determining the scaling function for the superdiffusive phase where the nonlinear fluctuating hydrodynamics \cite{Spohn} could be a promising direction.

\bibliography{cETH}

\end{document}